\documentclass[apjl]{emulateapj}
\usepackage{natbib}
\usepackage[usenames]{color}
\usepackage{ulem}
\newcommand{\be}{\begin{equation}}
\newcommand{\ee}{\end{equation}}
\newcommand{\bea}{\begin{eqnarray}}
\newcommand{\eea}{\end{eqnarray}}

\newcommand{\mpci}{\mbox{$h$\,Mpc$^{-1}$}}
 
\begin{document}
\title{Baryon Oscillations and Consistency Tests for 
Photometrically-Determined Redshifts of Very Faint Galaxies}
\author{Hu Zhan and Lloyd Knox}
\shortauthors{Zhan \& Knox}
\shorttitle{Photometric Redshift Consistency}
\affil{Department of Physics, University of California, Davis, CA 95616}
\email{zhan@physics.ucdavis.edu}
\email{lknox@physics.ucdavis.edu}
 
\begin{abstract}
Weak lensing surveys that can potentially place strong constraints 
on dark energy parameters can only do so if the source redshift 
means and error distributions are very well known.  We investigate 
prospects for controlling errors in these quantities by exploiting 
their influence on the power spectra of the galaxies. Although, 
from the galaxy power spectra alone, sufficiently precise and 
simultaneous determination of  redshift biases and variances is not 
possible, a strong consistency test is.  Given the redshift error 
rms, galaxy power spectra can be used to determine the mean 
redshift of a group of galaxies to subpercent accuracy.  Although 
galaxy power spectra cannot be used to determine the redshift error 
rms, they can be used to determine this rms divided by the Hubble 
parameter, a quantity that may be even more valuable for 
interpretation of cosmic shear data than the rms itself.  We also 
show that galaxy power spectra, due to the baryonic acoustic 
oscillations, can potentially lead to constraints on dark energy that are 
competitive with those due to the cosmic shear power spectra from the same survey.
\end{abstract}
 
 
\section{Introduction} \label{sec:intr}
Prospects of cosmic shear for cosmology \citep{hu99, hutegmark99, 
hu02, huterer02, heavens03, refregier03, benabed04, ishak04, sk04, 
takadajain04, takadawhite04, knox05b} have inspired many
on-going and future weak lensing surveys such as the Deep Lens 
Survey\footnote{See \url{http://dls.physics.ucdavis.edu.}},
Canada France Hawaii Telescope Legacy 
Survey\footnote{See \url{http://www.cfht.hawaii.edu/Science/CFHLS/.}}, 
and the Large Synoptic Survey 
Telescope\footnote{See \url{http://www.lsst.org.}} 
\citep[LSST,][]{tyson03}.

The statistical properties of cosmic shear maps depend on the
distances to the lenses and sources.  With the redshifts of
the source galaxies determined, the data can thus be
used to constrain dark energy parameters via their influence
on the distance-redshift relation \citep*{simpson05,knox05b}.  
Because the source galaxies
are faint and numerous their redshifts will be estimated from
multi-band imaging data (photometric redshifts) rather
than from spectroscopic data (spectroscopic redshifts).  For redshift
uncertainties not to degrade the constraints on dark energy parameters
one must know the rms photometric redshift error and any biases very
accurately (\citealt{bernstein04}; \citealt{huterer05b};
\citealt{ishak05}; \citealt*{ma05}).

Photometric redshifts are traditionally calibrated with spectroscopy
of a subset of the imaged galaxies. One needs a very large and fair
sample though in order to achieve a sufficiently accurate calibration.
For example, to reduce the uncertainties of the rms photometric
redshift error to 1\%, a minimum of 5000
spectra are required in each redshift bin and ideally for each
spectral type of galaxy. This is achievable at low redshift, but it becomes
prohibitively expensive toward high redshifts for the magnitude limits
of the deepest planned surveys.  Moreover, photometric
redshift errors are not Gaussian; the distributions tend to have long
tails.  This means that the rms value may not be
sufficient for characterizing photometric errors and that one needs
an even larger spectroscopic subsample to map the full error
distribution. Thus, it remains a challenge for very deep surveys to
calibrate their photometric redshifts with sufficient accuracy.

%\addt{Hu, I removed a paragraph here because the many-band approach
%is Tony's idea and we don't really do anything with it in the paper.
%I've tried to clean up any subsequent references to it.  It's still
%in ms3.tex.}

However this challenge is met,
we expect that independent cross checks capable of revealing
inconsistencies with their nominal error distributions.
will be very valuable.  The clustering properties of galaxies
in redshift space provide us with such an opportunity.
Photometric redshift errors, like peculiar velocities, alter
the galaxy power spectrum in redshift space. Specifically, they
strongly suppress the power spectrum along the line of sight. 
The amount of suppression is exponentially sensitive to the rms
photometric redshift error (e.g.~\citealt{se03}, hereafter SE03;
\citealt{zhan05}). Thus, one can use the three-dimensional galaxy 
power spectrum to quantify photometric redshift errors and, 
consequently, prevent weak lensing statistics from severe degradation. 

The clustering properties of the galaxies are also dark energy
probes in their own right.
The features in the galaxy power spectrum (the broadband shape and 
baryon acoustic oscillations)
\citep{py70, be84, h89, hs96, cpp05, ezh05, huetsi05} can be used 
as CMB-calibrated standard rulers 
for determining the angular-diameter distance $D(z)$ and 
constraining dark energy (\citealt*{eht98}; \citealt{bg03}; 
\citealt{hh03}; \citealt{l03}; SE03; \citealt{abf05}). 

Several papers have studied the prospects for measuring
baryon acoustic oscillations from photometric surveys
\citep*[SE03;][]{bb04, djt04, gb05, linder05}. The main advantage 
of a photometric redshift survey, is the wide coverage, which 
reduces the sample variance error, and deep photometry, which 
leads to more galaxies and therefore lower shot noise. 

The challenges for large photometric redshift surveys include
redshift errors, dust extinction, galaxy bias, 
redshift distortion and nonlinear evolution \citep{zhan05}.
These non-idealities do not produce strong oscillating features in 
the power spectrum (similar to the baryon oscillations) and can be 
controlled, despite some degradation to the measurements
\citep{se05, white05, zhan05}.

Ideally, the clustering properties of the galaxies would be adequate
to simultaneously constrain dark energy and the error distribution
of the photometric redshift errors.  As we show below though, 
such a self-calibration does not achieve a sufficiently accurate
reconstruction of the photometric redshift error distribution. 
However, the clustering properties of the galaxies {\em can} provide 
a valuable, though somewhat model-dependent, consistency test.

In section 2 we present our method for forecasting constraints on
cosmological parameters as well as parameters of the redshift error
distribution.  In Section 3 we show how well the rms photometric
redshift error can be estimated given the knowledge of the Hubble
parameter $H(z)$, and vice versa. We then demonstrate that mean 
redshifts can be accurately determined from the same data.
In Section 4 we forecast constraints on dark energy
from the broadband shape and baryon oscillations in the galaxies of 
the LSST survey. Since the constraints from the former alone are 
sub-dominant to those from the latter (SE03), hereafter we only 
refer to baryon oscillations even though the broadband shape is 
always included in our analysis.
In Section 5 we discuss these results 
and conclude.
 
\section{Method} \label{sec:method}
SE03 developed a two-stage Fisher matrix analysis to forecast 
errors on dark energy equation of state (EOS) parameters from 
baryon oscillations. 
We adopt their method with improvements and constrain photometric 
redshift errors at the same time. 

In the first stage of the SE03 analysis $H_i$ and $D_i$, where
$H_i \equiv H(z_i)$, $D_i \equiv D(z_i)$, and 
$z_i$ is the mean redshift of the $i$th bin of galaxies, are treated
as free parameters and their Fisher matrix is calculated.  In the
second stage, constraints on the matter density $\omega_{\rm m}$, 
the comoving angular diameter distance to the last scattering 
surface $D_{\rm CMB}$, $H_i$ and $D_i$ are converted
to constraints on dark energy parameters.  
The point of dividing the analysis into two stages is to allow
one to better understand how the dark energy constraints are arising.  

\subsection{Observed Galaxy Power Spectrum}

Using a fiducial cosmological model as reference, we can 
write the observed galaxy power spectrum as
\bea \label{eq:Pg}
P_{\rm g}(k_{{\rm f}\perp},k_{{\rm f}\parallel}) 
=&&\frac{D_{\rm f}^2 H}{D^2 H_{\rm f}} 
\left(1+\beta\frac{k^2_\parallel}{k^2_\perp + k^2_\parallel}
\right)^2  \\ && \times 
\big|\tilde W(c^2 \sigma_z^2 k_\parallel^2 / H^2)\big|^2
b^2 G^2 P(k) +P_{\rm s}, \nonumber
\eea
where $\beta$ %$\simeq \Omega_{\rm m}^{0.6}(z)/b$ \citep{llp91} 
is the linear redshift distortion parameter \citep{k87}, 
$\sigma_z = \sigma_{z0}(1+z)$ is 
the rms photometric redshift error, $b$ is the galaxy clustering 
bias, $G$ is the linear growth function, $P(k)$ is the matter power 
spectrum at $z = 0$, and $P_{\rm s}$ is the shot noise.
We have suppressed the argument $z$ in functions $D$, $D_{\rm f}$,
$H$, $H_{\rm f}$, $\beta$, 
$\sigma_z$, $b$, and $G$ for convenience. The true 
wavenumbers $k_\perp$ and $k_\parallel$ are related to the fiducial
ones by
\be
k_\perp = k_{\rm f\perp} D_{\rm f} / D \mbox{\, and \,}
k_\parallel = k_{\rm f\parallel} H / H_{\rm f}.
\ee

The window function, $\tilde W$, in Eq.~(\ref{eq:Pg}), 
is the Fourier transform in the radial direction of the real-space window function,
$W$, which is the distribution of galaxies at true distance $r$
given their estimated distance of $\bar r$.  
With the assumption that this distribution has the form
\be
P(r | \bar r) = W\left(\left(r-\bar r\right)/\sigma_r\right)
\ee
the modulus of the Fourier transform only depends on $k_\parallel \sigma_r =
k_\parallel c\sigma_z/H$.  If we further assumed a normal distribution
then $|\tilde W|^2 = e^{-c^2 \sigma_z^2 k_\parallel^2 / H^2}$ and
Eq.~(\ref{eq:Pg}) would be the same as in SE03.

In all our calculations, for simplicity, we do assume a normal distribution
for the redshift errors.  But in general redshift errors are significantly
non-Gaussian.  Typically
photometric redshift errors have long tails due to the fraction of redshift
assignments that fail catastrophically.  However, as can be seen from 
Eq.~(\ref{eq:Pg}) any non-Gaussianity has its impact on the observed galaxy 
power spectrum and, in principle, by assuming the isotropy of $P(k)$, 
one can reconstruct the function $\tilde W$ from the anisotropy of the 
observed galaxy power spectrum.  Of course, with greater freedom allowed 
in the functional form of the redshift error distribution, the quality of 
the reconstruction will be weakened.  

\citet{s04} points out that the \citet{k87} formula for the linear 
redshift distortion [used in Eq.~(\ref{eq:Pg})] may not 
be sufficiently accurate for the precision measurements we consider here. 
Fortunately, the effect of redshift distortion can be readily calibrated 
through $N$-body simulations. Meanwhile, one can also quantify the window 
function $W$ by assigning realistic redshift errors to mock galaxies 
drawn from the simulations. Precisely calibrated matter power spectra
are, in fact, required for interpreting weak lensing data \citep*{white04,
zhan04, huterer05a, hagan05}. As such, calibrating the effects of redshift
distortion and photometric redshift errors can be achieved from the same
set of simulations without much difficulty.

Although we treat them as constant across each redshift bin,
all the functions ($D$, $D_{\rm f}$, $H$, $H_{\rm f}$, 
$\beta$, $\sigma_z$, $b$, and $G$) in equation (\ref{eq:Pg}),
as well as the survey selection function, can 
change considerably within a single redshift bin. 
As such, the observed galaxy power spectrum in a redshift bin 
becomes a convolution of equation (\ref{eq:Pg}) with another 
window function that accounts for the radial evolution 
\citep[e.g., as treated in][]{zhan05}. We avoid this complication
for simplicity, although the appropriate treatment is straightforward
as long as the true evolution within each bin is known, or can
be parameterized.

The sub-bin evolution we are most concerned with is that of the 
clustering bias.  The variation will be caused in part by 
any redshift dependence of the selection function.  Even if one 
were selecting the same distribution of galaxy types at 
each redshift, bias would change due to the intrinsic evolution of 
the population.  Most importantly, redshift 
errors might actually be correlated with bias fluctuations
since both can be caused by spectral type errors 
\citep{padmanabhan05b}.  We do not yet know the magnitude
of these effects and the extent to which they can be controlled.  

\subsection{Fisher Matrix} \label{sec:fish}
For the stage I analysis, our complete parameterization of the observables
includes more than just $H_i$ and $D_i$.  There are
four other parameters in each redshift bin: $\ln(G_ib_i)$, $\ln\beta_i$, 
$P_{{\rm s},i}$ and $\ln{\sigma_{z,i}}$, where $i$ refers to the $i$th 
redshift bin, and seven redshift-independent parameters.  
We do not include the redshift bias parameters, $\delta z_i$, at
this stage because they have no influence on the observed power spectra
except through $H_i$ and $D_i$, which are independent parameters in stage I.
They will be included in stage II. For brevity, we suppress the subscript
$i$ in what follows.

The seven redshift-independent parameters are the matter density 
$\omega_{\rm m}$, the baryon density $\omega_{\rm b}$, 
the redshift of reionization $z_{\rm rei}$, the primordial helium mass
fraction $y_{\rm p}$, the spectral tilt $n_{\rm s}$, the normalization 
of the primordial potential power spectrum $k_0^3P_{\Phi 0}$, and 
the angular size of the sound horizon at the last 
scattering surface $\theta_{\rm s}$. 

There are three major difference between our choice of parameters and that 
of SE03.  First, we do not include the matter fraction 
$\Omega_{\rm m}$ because at fixed $\theta_{\rm s}$ and $\omega_{\rm m}$
it affects neither the CMB observables nor the 
matter power spectrum.  Second, 
we use $\ln Gb$ instead of $\ln G$ because the galaxy clustering
bias is degenerate with the growth function. And third, we include
additional parameters $\ln \sigma_z$ (stage I) and $\delta z$ 
(stage II) in each redshift bin. We also replace $D_{\rm CMB}$ in SE03 
with $\theta_{\rm s}$ for convenience, which is not critical.

As the cosmic density field evolves, nonlinear
effects become important on larger and larger scales.
To prevent these effects from contaminating the measurements of
baryon oscillations, we only use the Fourier modes at
$k < k_{\rm max}$ in our analysis, where $k_{\rm max}$ ranges from
0.12 to 0.53 \mpci{} between $z = 0.31$ and $2.66$ (see
Table~\ref{tab:err}). Evidence from the 2-degree Field Galaxy Redshift
Survey and the Sloan Digital Sky Survey suggests that the
galaxy clustering bias at $z \sim 0.1$, despite some dependence on
luminosity, is scale-independent on scales $k$ less than a few tenths
\mpci{} \citep[e.g.][]{verde02, tegmark04}.
Hence, the upper bound in wavenumber also ensures that we can model
the galaxy bias as a time-evolving but scale-independent quantity.
For simplicity, our fiducial model has a bias $b(z) = 1 + 0.84 \, z$ 
for all galaxies.  It is straightforward to model the bias for 
each sub-sample of galaxies separately. In fact, it is beneficial to
divide the galaxies into sub-samples, especially at low redshift 
where galaxy number density is high, because one can control 
photometric redshift errors much better for a homogeneous sample of
galaxies \citep[e.g.][]{pbs05}.

We assume that the primordial potential power spectrum
$k^3P_{\Phi}^i(k) = k_0^3P_{\Phi 0} (k/k_0)^{n_{\rm s}-1}$.
Our fiducial model is a low density and flat cold dark matter
universe with a cosmological constant. It has
$(\omega_{\rm m}, \omega_{\rm b}, z_{\rm rei},
y_{\rm p}, n_{\rm s}, k_0^3P_{\Phi 0}, \theta_{\rm s}) =
(0.146, 0.021, 6.3, 0.24, 1.0, 6.4 \times 10^{-11}, 0.60 \ {\rm deg})$
and the reduced Hubble constant $h = 0.655$.

We calculate the Fisher matrix using the approximation \citep{t97}
\be \label{eq:Fij}
F_{ij} = \int \frac{1}{2} 
\frac{\partial \ln P(\mathbf{k}_{\rm f})}{\partial p_i} 
\frac{\partial \ln P(\mathbf{k}_{\rm f})}{\partial p_j} 
V_{\rm eff}(\mathbf{k}_{\rm f}) 
\frac{{\rm d} \mathbf{k}_{\rm f}}{(2 \pi)^3},
\ee
where $p_i$ is the $i$th parameter, 
\be \label{eq:Veff} 
V_{\rm eff}(\mathbf{k}) = \int \left [ 
\frac{{n}(\mathbf{r})P_{\rm g}(\mathbf{k})}
{{n}(\mathbf{r})P_{\rm g}(\mathbf{k})+1}
\right ]^2 {\rm d} \mathbf{r}
\ee
is the effective volume of the survey \citep{fkp94}, and 
$n(\mathbf{r})$ is the galaxy number density. The marginalized
error of the $i$th parameter is given by 
$[(F+F_{\rm P})^{-1}]^{1/2}_{ii}$.  Here, $F_{\rm P}$ is for any prior
information.  If we already know, prior to examining
the data, the value of parameter $p_i$ to within $\pm \sigma_{\rm P}(p_i)$ then
$(F_{\rm P})_{ii} = 1/\sigma_{\rm P}^2(p_i)$.  

In the second stage, our ``observables" are 
$\omega_{\rm m}$, $\theta_{\rm s}$, $D$'s, and $H$'s 
with their covariance matrix as calculated from the stage I
Fisher matrix, marginalizing over all the other parameters.
From these observables, and their covariance matrix, we forecast
how well one can determine the following parameters:
the photometric redshift biases $\delta z$, 
$\omega_{\rm m}$, $\Omega_{\rm DE}$, 
$w_0$, and $w_{\rm a}$, where the dark energy EOS is parametrized as
$w(z) = w_0 + w_{\rm a}[1 - (1 + z)^{-1}]$.
To explicitly define the redshift bias parameters we write
the probability that a galaxy with photometric 
redshift estimate $z_{\rm p}$ has a true redshift in between 
$z$ and $z+{\rm d}z$ as
\be
{\rm d}P(z)  =  \frac{1}{\sqrt{2\pi}\,\sigma_z}
\exp\left[-\frac{(z +  \delta z - z_{\rm p})^2}
{2\sigma_z^2}\right]  {\rm d}z.
\ee

\begin{deluxetable*}{l c c c c c c c }
\tablewidth{0pt}
\tablecaption{Marginalized Errors of Selected Stage I Parameters \\
with Complete Self-Calibration from LSST
\label{tab:err}}
\tablehead{\colhead{\phd $z$} & \colhead{$k_{\rm max}$}  & 
\colhead{\phd $\sigma(\ln D)$\tablenotemark{a}} & 
\colhead{$\sigma(\ln H)$\tablenotemark{a}} & 
\colhead{$\sigma(\ln H)$\tablenotemark{b}} & 
\colhead{$\sigma(\ln \sigma_z/H)$\tablenotemark{a,c}} &
\colhead{\phd $\sigma(\ln \beta)$\tablenotemark{a}} & 
\colhead{\phd $\sigma(\ln Gb)$\tablenotemark{a}} \\
\cline{3-8} 
\colhead{} & \colhead{($h$\,Mpc$^{-1}$)} & 
\multicolumn{6}{c}{($\times 100$)} }
\startdata
0.31 & 0.12 & \phn 2.9 & 27 & 1.1 & 0.37 & \phd 24 & \phd 13 \\
0.55 & 0.14 & \phn 1.7 & 19 & 1.0 & 0.21 & \phd 17 &     9.5 \\
0.84 & 0.17 &     0.75 & 14 & 1.0 & 0.14 & \phd 13 &     7.1 \\
1.18 & 0.21 &     0.57 & 12 & 1.0 & 0.11 & \phd 11 &     6.2 \\
1.59 & 0.28 &     0.43 & 10 & 1.0 & 0.10 &     9.4 &     5.2 \\
2.08 & 0.38 &     0.39 & 11 & 1.0 & 0.11 & \phd 10 &     5.5 \\
2.66 & 0.53 &     0.43 & 12 & 1.0 & 0.15 & \phd 13 &     6.4 
\enddata
\tablenotetext{a}{
No prior is taken for the rms photometric redshift error or the 
Hubble parameter. The constraints are poorer at low redshift due 
to small volumes and low $k_{\rm max}$ and then are poorer at high 
redshift due to low galaxy number densities.}
\tablenotetext{b}{
We assume $\sigma_{\rm P}(\ln \sigma_z) = 0.01$ for this column. 
The prior sets the precision of $H$, 
reduces uncertainties of $\beta$ and $Gb$ significantly, and
renders little improvement to other parameters. }
\tablenotetext{c}{
For this column, the parameter $\ln \sigma_z$ in the Fisher
matrix is replaced by $\ln \sigma_z/H$. With this 
parametrization, the constraints on other parameters change 
slightly. }
\end{deluxetable*}

\subsection{Tests} \label{sec:test}
We test our procedures using the baseline surveys of SE03. In 
order to compare with their results, we do not include photometric
redshift parameters and do not differentiate the photometric 
redshift term $e^{-c^2 \sigma_z^2 k_\parallel^2 / H^2}$ with
respect to $H$ in the Fisher matrix. We also adopt SE03 
parametrization of the dark energy EOS in this test. The resulting 
uncertainties on $D$ and $H$ match those in SE03 within a factor 
of $1.2$ ($1.45$) for the spectroscopic (photometric) baseline 
survey. Our error forecasts on dark energy parameters are 
$30\%$--$50\%$ larger than those in SE03 because of the 
larger errors we obtain for $D$ and $H$.

If we include the derivative of the photometric redshift term 
$e^{-c^2 \sigma_z^2 k_\parallel^2 / H^2}$ with respect to 
$H$ in the Fisher matrix, then the constraints on $H$ and dark
energy EOS parameters are nearly as good as those from the 
spectroscopic baseline survey. This is because the shape of the 
observed galaxy power spectrum is exponentially sensitive to the 
Hubble parameter. Such sensitivity imposes a strong constraint on 
$H$ and subsequently on dark energy EOS parameters. However,
the strong constraint is actually on the comoving length scale of 
the rms photometric redshift error $c \sigma_z/H$. In 
reality, the uncertainty of $\sigma_z$ will degrade the 
constraint on $H$. In fact, if we also include $\ln \sigma_z$ 
in the parameter set, the constraints become as 
weak as those from the photometric baseline survey in the previous 
paragraph. Thus, we conclude that the photometric 
results in SE03 are equivalent to having an imprecise prior
on $\sigma_z$; i.e., letting $\sigma_z$ float freely. 

\section{Results for Redshift Parameters}
\label{sec:photoz}

We assume that through some external calibration process we
have estimates of both the distribution of errors about the mean redshift
as well as the mean redshift.  A central point of our paper
is that one can use the clustering properties of the galaxies 
themselves to test these
externally-derived mean redshifts.  To explain how it works, let
us assume that the distribution of errors is Gaussian with variance
$\sigma_z^2$.  The redshift errors lead to a suppression of the fluctuation 
power of radial modes by an amount $\exp(-c^2\sigma_z^2k_\parallel^2/H^2)$.
With 
$\sigma_z$ known, one can thus determine $H$.  Analyzing the power spectra
of a sequence of galaxy groupings, each with mean redshifts, $z_{\rm m}$,
we thus know $H(z_{\rm m} - \delta z)$ for each group of galaxies.
% where $\delta z$ is the bias
%in our estimation of the mean redshift for the $i$th group of galaxies.
From the standard ruler provided by the baryon oscillations, we
can also determine $D(z_{\rm m} - \delta z)$.  With
some assumptions about the smoothness of $H$, and using the fact that
$D = c \int dz/H$, these values of $D$ and $H$ will not be consistent with
each other for all possible values of $z_{\rm m} - \delta z$.  
Thus we can determine these mean redshifts and check to see if they are 
consistent with the mean redshifts inferred from the external calibration.

Even if we do not use the external calibration to tell us $\sigma_z$, 
from the galaxy power spectra we can still determine the source 
distance rms $\sigma_r = c \sigma_z/H$ to much better than $1\%$ 
(see Table~\ref{tab:err}).  Fortunately, this is
exactly the quantity we want for interpretation of weak lensing
power spectra, since the weak lensing power spectrum for sources
with mean redshift $z$ depends on how the sources are distributed
in real space, not redshift space.

In this section we investigate quantitatively how well one can measure
$\sigma_z$, given $H$ (and vice-versa), perform the redshift
bias consistency test and determine $\sigma_z/H$.   
For specificity, we adopt for this section and the next a model of the 
proposed survey by LSST as our fiducial survey.  We assume a survey area 
of $23,000$ square degrees with galaxy distribution 
\be
n(z) = 640 z^2 e^{-z / 0.35}\ \mbox{arcmin}^{-2},
\ee
which corresponds to a projected number density of 55 
\mbox{arcmin$^{-2}$}. We use $\sigma_{z0} = 0.04$ 
as our fiducial rms photometric redshift error.
The survey is divided into 7 redshift bins from $z = 0.2$ to 
$3$ each with roughly the same comoving radial extent. 
The mean redshifts of the redshift bins are listed in 
Table~\ref{tab:err}.

For the CMB we assume {\it Planck} as treated in \citet*{kaplinghat03}. 
{\it Planck} observes at 9 different frequency channels from 30 to 850 
GHz. To crudely model the effects of foregrounds we assume the
temperature maps at 100, 143 and 217 GHz can be perfectly cleaned
of foregrounds by use of the other channels and for polarization
we keep just the 143 and 217 GHz channels.  We assume this can
be done over 80\% of the sky up to $l_{\rm max} =
2000$ for temperature and $l_{\rm max} = 2500$ for polarization.

\subsection{rms Photometric Redshift Error and $H$} \label{sec:szh}

\begin{figure}
\centering
\includegraphics[width=72mm]{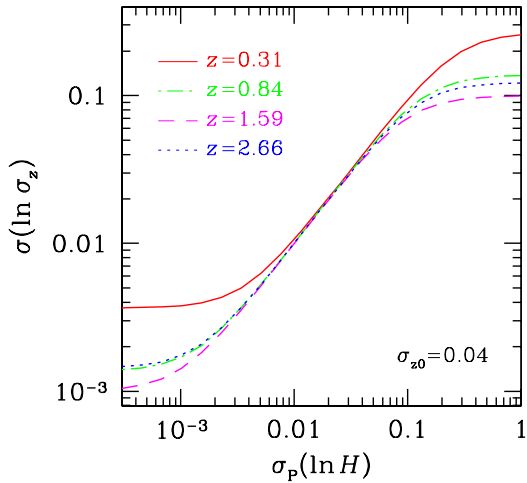}
\caption[f1]{Fractional errors on the rms photometric redshift 
error $\sigma(\ln \sigma_z)$ as a function of the prior on the 
Hubble parameter $\sigma_{\rm P}(\ln H)$ in four of the seven redshift bins 
assuming the fiducial LSST survey.  On the left side, the 
prior is better than the constraint of the combination 
$\sigma_z/H$ (see Table~\ref{tab:err}), so that it can no longer 
improve the precision 
of $\sigma_z$. Whereas, on the right side, the prior is not 
useful because the survey constrains $H$ better than the prior, i.e.
it reaches self-calibration.
In the middle, one has $\sigma(\ln \sigma_z) = \sigma_{\rm P}(\ln H)$ as 
expected. The uncertainty of $H$ as a function of the prior on
$\sigma_z$ look identical to this figure except the x-axis
label becomes $\sigma_{\rm P}(\ln \sigma_z)$ and the y-axis
label $\sigma(\ln H)$.
\label{fig:szh}}
\end{figure}

We first implement the full procedure described in Section 
\S~\ref{sec:method} with a prior on the Hubble parameter 
$\sigma_{\rm P}( \ln H)$. For a simple demonstration, we apply the same
prior to all redshift bins. The results for four of the
redshift bins are shown in 
Fig.~\ref{fig:szh}.  The other three are suppressed for clarity.  
As expected, the prior 
on $H$ sets directly the uncertainty of the rms photometric 
redshift error, i.e.~$\sigma(\ln \sigma_z) = \sigma_{\rm P}(\ln H)$, 
over a wide range. One is limited by the uncertainty of the 
combination $\sigma_z/H$ when the prior
$\sigma_{\rm P}( \ln H)$ is stronger than the constraint 
$\sigma[\ln (\sigma_z/H)]$, so that $\sigma_{\rm P}( \ln H)$
can no longer improve the precision of $\sigma_z$. 
On the other hand, $\sigma(\ln \sigma_z)$ does not degrade 
further when the prior is worse than what can be determined from 
the survey itself. Hence, the 
$\sigma(\ln \sigma_z)$--$\sigma_{\rm P}( \ln H)$
curves in Fig.~\ref{fig:szh} flatten on both sides.

A prior on the Hubble parameter could potentially come from a 
spectroscopic galaxy
survey such as that planned with the Kilo-Aperture Optical 
Spectrograph\footnote{See http://www.noao.edu/kaos/.}.  KAOS will
be able to survey enough volume to determine the Hubble 
parameter at $z \sim 1$ and 3 to a few percent from baryon
oscillations (SE03). 

On the other hand, we may take a prior on $\sigma_z$ from a 
spectroscopic calibration of the photometric redshifts. We have 
thus also calculated the effect of a prior on the rms photometric 
redshift error $\sigma_{\rm P}(\ln \sigma_z)$ on our ability
to determine $H$.  Again, the same prior was taken for the
seven redshift bins.  The results are almost identical 
to those in Fig.~\ref{fig:szh} but with x-axis label becoming 
$\sigma_{\rm P}(\ln \sigma_z)$ and the y-axis label 
$\sigma(\ln H)$, because $\sigma_z$ 
and $H$ are degenerate in the photometric suppression
$e^{-c^2 \sigma_z^2 k_\parallel^2 / H^2}$. 

%The errors in the angular diameter distances decrease from $2.9\%$ 
%in the lowest redshift bin to $0.43\%$ in the highest redshift bin
%(see Table~\ref{tab:err}).
%They are reduced only very slightly by applying the prior 
%$\sigma_{\rm P}(\ln \sigma_z) = 0.01$.

\subsection{Photometric Redshift Bias}

In stage II of the Fisher matrix analysis, we include
redshift biases $\delta z$. 
These redshift biases will cause inconsistencies between 
the angular diameter distance and Hubble parameter in a given 
cosmological model, given some assumptions about the smoothness
of $H$.  In our analysis, those smoothness assumptions are
provided implicitly by modeling the dark energy density as
a smoothly varying function, controlled by EOS parameters $w_0$ and
$w_{\rm a}$. 

\begin{figure}
\centering
\includegraphics[width=72mm]{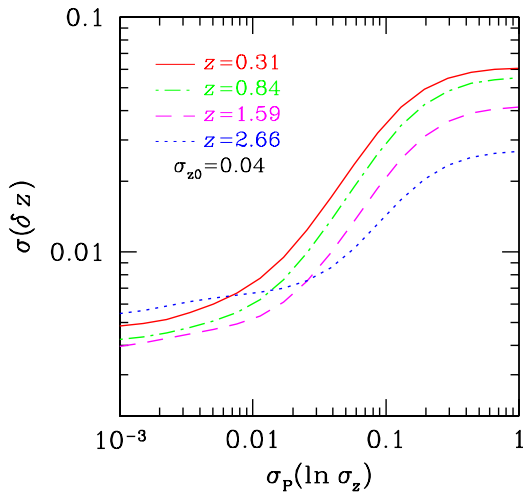}
\caption[f2]{Forecasted errors on $\delta z$ given varying priors 
on $\sigma_z$. At low $\sigma_{\rm P}(\ln{\sigma_z})$ the quality 
of the constraint as a function of $z$ peaks at
intermediate redshifts, due to the similar behavior for the
constraints on $D$ and $H$ (see Table~\ref{tab:err}).  
%The constraints on $D$ and $H$ are poor
%at low redshift due to small volumes and low $k_{\rm max}$ and then
%are poor at high redshift due to low galaxy number densities.
\label{fig:zbias}}
\end{figure}

The resulting constraints on the photometric redshift biases are
shown in Fig.~\ref{fig:zbias} as a function of the input prior
$\sigma_{\rm P}(\ln \sigma_{\rm z})$, which is the same in all 
redshift bins. With $\sigma_{\rm P}(\ln \sigma_{\rm z}) \sim 0.01$
the biases are determined to better than 0.01.
For the 4,000 sq. degree cosmic shear survey considered by
\citet{ma05}, these bias constraints are sufficient to keep
the degradation in $w_0$ and $w_{\rm a}$ errors at the 20\% and 50\%
levels respectively.  However, because the larger LSST cosmic shear
survey can achieve stronger constraints on $w_0$ and $w_{\rm a}$ (given perfect
redshift information), the degradation in $w_0$ and $w_{\rm a}$ in this
case is worse:  factors of 2.5 and 3 respectively\footnote{Zhaoming
Ma, private communication.}.  This degradation is much better though
than the factor of ~10 degradation in $w_0$ and $w_{\rm a}$ that occur
without any prior on the redshift biases.

At low $\sigma_{\rm P}(\ln{\sigma_z})$ the quality of the constraint 
as a function of $z$ peaks at
intermediate redshifts, due to the similar behavior for the
constraints on $D$ and $H$ (see Table~\ref{tab:err}).  

\section{Dark Energy Constraints from LSST Baryon Acoustic
Oscillations} \label{sec:lsst}

While the constraints on $\sigma_z$ and $\delta z$ from galaxy 
power spectra discussed in Section~\ref{sec:photoz} will be useful
for extracting dark energy constraints from cosmic shear data, we
also explore how well dark energy information can be extracted
from the galaxy power spectra themselves.  

\begin{figure}
\centering
\includegraphics[width=72mm]{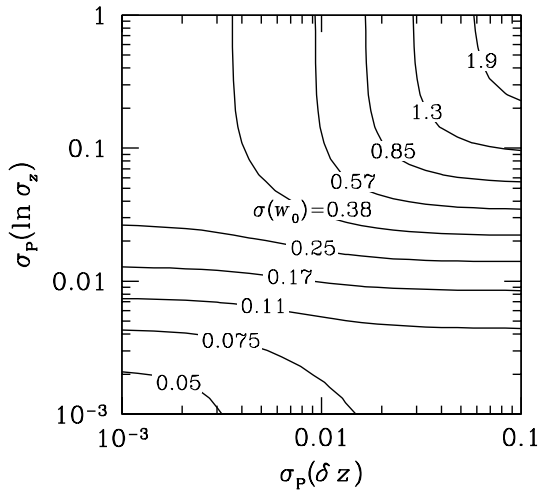}
\caption[f3]{Contours of constant forecasted errors in $w_0$ estimated 
from galaxy clustering in the LSST survey as priors on 
$\delta z$ and $\sigma_z$ are varied.  The same priors 
are applied to all redshift bins for simplicity.  
\label{fig:contw0}}
\end{figure}

\begin{figure}
\centering
\includegraphics[width=72mm]{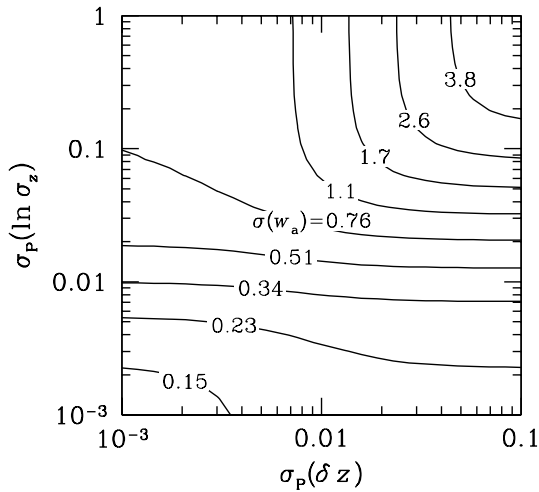}
\caption[f4]{As in Fig.~\ref{fig:contw0}, but the contours 
are for constant error in $w_{\rm a}$.
\label{fig:contwa}}
\end{figure}

In Figures ~\ref{fig:contw0} and ~\ref{fig:contwa} we see contours 
of constant forecasted errors in $w_0$ and $w_{\rm a}$ as the 
redshift priors are varied.  The same priors 
are applied to all redshift bins for simplicity. As expected,
the galaxy power spectrum alone provides for only a very weak 
self-calibration; as the prior on rms redshift errors is relaxed 
to $100\%$ and that on redshift biases to $0.1$, 
the errors degrade by factors of 20
to 40.  Also as expected, we do see though that with a strong prior
on $\sigma_z$, one does not need a prior on the $\delta z$ parameters.  

From Fisher matrix calculations like those in Section~\ref{sec:photoz} 
we find that the Hubble parameter does not significantly improve the 
constraints on dark energy EOS parameters unless it is determined with
precision comparable to or better than those of the angular 
diameter distance.  This means that one must achieve
$\sigma(\ln H) \lesssim 1\%$, which is possible, as discussed in 
Section~\ref{sec:szh}, with a prior of 
$\sigma_{\rm P}(\ln \sigma_z) \lesssim 1\%$.

\begin{figure}
\centering
\includegraphics[width=72mm]{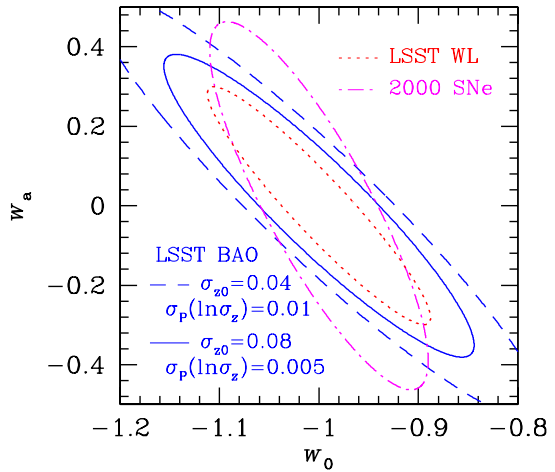}
\caption[f5]{Marginalized $1\sigma$ error contours in the 
$w_0$--$w_{\rm a}$ plane. The contours for LSST baryon 
acoustic oscillations are shown for 
$[\sigma_z, \sigma_{\rm P}(\ln \sigma_z), \sigma_{\rm P}(\delta z)]
= (0.04, 0.01, 0.01)$ (dashed line) and $(0.08, 0.005, 0.08)$ 
(solid line). For comparison, we include the 
error ellipses from the power spectra of LSST tomographic cosmic shear
maps (dotted line) and 2000 supernovae as might be observed
by the Joint Dark Energy Mission 
\citep[dash-dotted line,][]{knox05a}.
Note that the weak lensing constraints, unlike the baryon oscillation
constraints, do not included the effects of redshift error distribution
uncertainties.
\label{fig:lsstw0wa}}
\end{figure}

We show LSST baryon oscillation constraints on $w_0$ and 
$w_{\rm a}$ in Fig.~\ref{fig:lsstw0wa} for two cases:
$[\sigma_z, \sigma_{\rm P}(\ln \sigma_z), \sigma_{\rm P}(\delta z)]
= (0.04, 0.01, 0.01)$ (dashed line) and 
$(0.08, 0.005, 0.08)$ (solid line). 
By reducing $\sigma_{\rm P}(\ln \sigma_z)$ from $0.01$ to $0.005$
one tightens the constraints considerably, even though the 
rms photometric redshift error is increased. This example 
demonstrates that a large rms 
photometric redshift error can be tolerated as long as it is 
known accurately. 

\begin{figure}
\centering
\includegraphics[width=72mm]{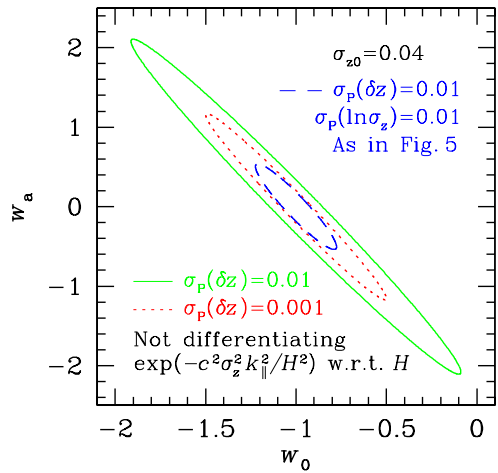}
\caption[f6]{As in Fig.~\ref{fig:lsstw0wa}, but the two larger 
error contours are obtained without differentiating the photometric
suppression term $\exp(-c^2 \sigma_z^2 k_\parallel^2 / H^2)$ with 
respect to the Hubble parameter $H$ in the Fisher matrix in stage I. 
The rms photometric redshift error is assumed to be 
$\sigma_z = 0.04(1+z)$, and it is not included as parameters. 
In stage II, the same prior 
is applied to the photometric redshift bias in each redshift bin. 
The solid contour corresponds to $\sigma_{\rm P}(\delta z) = 0.01$,
and the dotted contour $\sigma_{\rm P}(\delta z) = 0.001$. 
The constraints do not improve for 
$\sigma_{\rm P}(\delta z) \lesssim 0.001$ or degrade much for
$\sigma_{\rm P}(\delta z) \gtrsim 0.008$. The dashed contour is 
from Fig.~\ref{fig:lsstw0wa} for 
$[\sigma_z, \sigma_{\rm P}(\ln \sigma_z), \sigma_{\rm P}(\delta z)]
= (0.04, 0.01, 0.01)$.
\label{fig:selimit}}
\end{figure}

We also show in Fig.~\ref{fig:lsstw0wa} the constraints from  
the power spectra of LSST tomographic cosmic shear maps (dotted line) 
and 2000 supernovae as might be observed by the Joint Dark 
Energy Mission \citep*[dash-dotted line,][]{knox05a}.  Note that
these weak lensing error forecasts are done in the limit of perfect
prior knowledge of the redshift error distribution parameters.  If we
were to use the priors on these parameters adopted for the baryon
oscillation calculations, the weak lensing constraints would become
slightly looser than the baryon oscillation constraints. 

One may attempt to combine all the three constraints together
to achieve higher precision.  However, further work is needed
to carefully account for the correlation between 
baryon oscillations and weak lensing statistics, because they 
overlap in both observational data and underlying density field
\citep[e.g.][]{hu04}.

As mentioned earlier, we cannot yet be completely sure that the sub-bin evolution
of galaxy bias can be controlled well enough to make an accurate
determination of $\tilde W(c\sigma_z k_\parallel/H)$, the suppression of 
power in the radial direction.  If we are not able to extract the 
information about the shape of the window function and, therefore,
$\sigma_z/H$, then even with a prior on the redshift
error distribution from external calibration our determination of $H(z)$
will be degraded as discussed in Section~\ref{sec:szh}.  We show in
Fig.~\ref{fig:selimit} how the error contours degrade in the limit
of no information extracted from the shape of the radial suppression.  
Specifically, for this calculation we do not differentiate the 
photometric suppression
term $e^{-c^2 \sigma_z^2 k_\parallel^2 / H^2}$ with respect to $H$
in stage I, and the parameters 
$\ln \sigma_z$ are removed.  In other words, this is the calculation
in the SE03 limit as discussed earlier.
Unlike in SE03 we include photometric redshift biases $\delta z$ in stage 
II to explore their effect on dark energy constraints. The error 
contour with $[\sigma_z, \sigma_{\rm P}(\ln \sigma_z), 
\sigma_{\rm P}(\delta z)]= (0.04, 0.01, 0.01)$ (dashed line) 
from Fig.~\ref{fig:lsstw0wa} is added for comparison. 

One sees in Fig.~\ref{fig:selimit} that 
the constraints from baryon oscillations in photometric redshift 
surveys are tightened tremendously by fully utilizing our 
knowledge of the photometric suppression to the galaxy power 
spectrum. Without extracting information from 
$e^{-c^2 \sigma_z^2 k_\parallel^2 / H^2}$, 
we obtain $[\sigma(\Omega_{\rm DE}), \sigma(w_0), \sigma(w_{\rm a})]
= (0.046, 0.33, 0.77)$ for $\sigma_{\rm P}(\delta z) = 0.001$ 
(dotted line) and $(0.084, 0.60, 1.4)$ for
$\sigma_{\rm P}(\delta z) = 0.01$ (solid line). These constraints
are as loose as those with very weak priors on $\ln \sigma_z$ in 
Figs.~\ref{fig:contw0} and \ref{fig:contwa}. They do not improve for 
$\sigma_{\rm P}(\delta z) \lesssim 0.001$ or degrade much for
$\sigma_{\rm P}(\delta z) \gtrsim 0.008$.

\section{Discussion and Conclusions} \label{sec:dis}

The statistical properties of weak lensing maps are sensitive
to the distances to the sources, and not to the redshifts of
the sources. Therefore, even without redshift information, 
weak lensing observations can be used to determine the source
distance distribution.  The redshift information is needed
only so the source distance distribution can inform us about
the distance-redshift relation.  However, this distance-redshift
relation is exactly what we want, because of its sensitivity
to dark energy, and therefore the redshift information is actually
terribly important.

We have pointed out that the three-dimensional power spectra
of the source galaxies, because of their sensitivity to redshift
errors, can be used to inform us about the source redshift 
distribution.  While the galaxy power spectra, on their own,
cannot give us all we need to know for interpretation of
tomographic cosmic shear data, they can be used to {\em supplement} 
the redshift information available from external calibration of the
photometric redshifts.  

We have demonstrated the supplemental information available
in two different ways.  First, if one uses prior information
(perhaps from external calibration)
about the variance of the redshift errors
in each of the redshift bins, then the mean redshifts of 
each of those redshift bins can be recovered.  They
can be recovered with sufficient precision to greatly limit 
the degradation in the errors on dark energy EOS parameters
reconstructed from cosmic shear data.  Second, without
any prior on the redshift error variance, the three-dimensional galaxy
power spectra can be used to determine very precisely the
source distance distribution which with the assumption of
a normal distribution is specified by the combination $\sigma_z/H$.  
The source distance distribution
will be very helpful for interpretation of cosmic shear data.
Although, for determining the distance-redshift relation (and
thereby the dark energy EOS parameters) we will still
require mean redshifts for each of the redshift bins, in this
case from some external calibration.

The galaxy power spectra, combined with external calibration
of the source redshift distributions, are also potentially powerful
probes of dark energy.
{\em If} the rms photometric redshift error is determined to better 
than $1\%$ through some external calibration
of an unbiased subsample, and one can
control sub redshift bin bias evolution then
one can significantly reduce errors on the Hubble parameter and, 
consequently, put tight constraints on the dark energy EOS 
parameters using baryon acoustic oscillations. For instance, with 
priors $\sigma_{\rm P}(\ln \sigma_z) = 0.005$ and 
$\sigma_{\rm P}(\delta z) \lesssim 0.004$, one can achieve 
$\sigma(\Omega_{\rm DE}) = 0.015$, $\sigma(w_0) = 0.095$, and 
$\sigma(w_{\rm a}) = 0.24$ with LSST. 

Achieving, in practice, the constraints on dark energy from baryon 
oscillations
and cosmic shear, that are possible in principle, will require very
accurate calibrations of photometric redshifts for very faint galaxies.
This is a significant challenge for the observational community.  The 
consistency tests we have described here may play an important role in 
meeting that challenge.

\acknowledgements
We thank Y.~S. Song for providing the CMB Fisher matrix. 
We thank D. Eisenstein, Z. Ma, N. Padmanabhan, and J.~A. Tyson for 
useful conversations and McGill 
University for their hospitality while some of this work was completed. 
This work was supported by the National Science Foundation under 
Grant No. 0307961 and NASA under grant No. NAG5-11098.

%\bibliographystyle{apj}
%\bibliography{ref}

\end{document}